\documentclass[prl,a4paper,superscriptaddress,twocolumn,showpacs,amsmath,amssymb,floatfix]{revtex4-2}

\usepackage{graphicx}
\usepackage{hyperref}

\usepackage{xcolor}

\definecolor{myblue}{HTML}{1F77B4}
\definecolor{myorange}{HTML}{FF7F0E}
\definecolor{myred}{HTML}{D62728}

\definecolor{colorA0}{HTML}{4C9ED9}    
\definecolor{colorA01}{HTML}{1F77B4}   
\definecolor{colorA03}{HTML}{FF7F0E}   
\definecolor{colorA05}{HTML}{D62728}   
  
\definecolor{blueprl}{HTML}{1f77b4}
\definecolor{orangeprl}{HTML}{ff7f0e}
\definecolor{greenprl}{HTML}{2ca02c}
\definecolor{redprl}{HTML}{d62728}
\definecolor{violetprl}{HTML}{9467bd}
\definecolor{brownprl}{HTML}{8c564b}

\begin{document}

\title{Small-world structure of quantum computer hardware}

\author{S. J. da Silva Junior}
\email[]{silvio\_jonas.da\_silva\_junior@umlp.fr}
\affiliation{Université Marie et Louis Pasteur, CNRS, Institut UTINAM (UMR 6213),
  équipe de physique théorique, Besançon, France}
\author{D. L. Shepelyansky}
\email[]{dima@irsamc.ups-tlse.fr}
\affiliation{Univ Toulouse, CNRS, Laboratoire de Physique Théorique, Toulouse, France}
\author{J. Lages}
\email[]{jose.lages@umlp.fr}
\affiliation{Université Marie et Louis Pasteur, CNRS, Institut UTINAM (UMR 6213),
  équipe de physique théorique, Besançon, France}

\date{August 13, 2026}

\begin{abstract}
		We show that the network of quantum register states of a quantum computer (QC), coupled through residual two-body interactions between qubits, exhibits small-world properties analogous to those of complex networks found in human society. The most probable Erd\H{o}s number between any two states is about 9, comparable to the six degrees of separation reported by Milgram for social networks. Using the {\AA}berg criterion, which retains only interactions exceeding the local energy spacing, we construct an effective ({\AA}berg) network and show that, above a critical coupling strength, this network percolates into a giant component spanning nearly the whole quantum register space. This percolation transition closely matches the onset of quantum chaos and dynamical thermalization established previously via costly exact diagonalization, while our approach extends the accessible system size up to $n_q=30$ qubits.
\end{abstract}


\maketitle

{\it Introduction.-} The small-world Nebraska experiment of Milgram \cite{milgram} showed that people in the US have, on average, six degrees of separation between each other. More recently, it was shown that the Facebook community of 721 million active users has only $N_E=4$ degrees of separation \cite{vigna}. The small-world properties have been actively studied for various networks, as reviewed, e.g. in \cite{dorogovtsev,newman}. The Erd\H{o}s number $N_E$ defines the minimum degree of separation, or number of links, between a hub node and any other target node. It follows the measure proposed by Paul Erd\H{o}s to count the distance between him and his co-authors via joint scientific articles (see, e.g. \cite{dorogovtsev,newman}). In this work we show that the concept of the small-world also works for a generic model of a Quantum Computer (QC) hardware characterized by low values of the Erd\H{o}s number $N_E$. The fundamental properties of QC are described in \cite{steane,chuang,deutsch}. The residual couplings between QC register states represent an effective undirected weighted network. The weights of the links are determined by the two-body residual couplings between qubits. Above the critical two-body interaction strength $J_c$, determined by the {\AA}berg criterion \cite{aberg1,aberg2,jacquod,nobel,mirlin1,mirlin2}, there is a melting of quantum non-interacting eigenstates with the emergence of quantum chaos and dynamical thermalization. 
For a QC hardware, above this quantum chaos border with inter-qubit couplings $J > J_c$, there is a dynamical thermalization over the global network of quantum register states belonging to the associated exponentially large Hilbert space \cite{georgeot1,georgeot2,benenti,frahm1}. In this Letter, we show that the {\AA}berg criterion describes a percolation transition throughout this network in agreement with the results for the quantum dynamical thermalization established through the computationally demanding diagonalization of QC eigenstates \cite{georgeot1,georgeot2,benenti,frahm1}. It is known that there is a certain similarity between the percolation transition in classical systems (see, e.g. \cite{percolation1,percolation2,percolation3}) and the Anderson-type delocalization \cite{anderson1958} on the Bethe lattice (see the review \cite{leticia} and references therein). Here, based on the {\AA}berg criterion, we describe the small-world properties of QC hardware and show that the percolation concept captures the transition to quantum chaos and thermalization.

\begin{figure}[t]
\begin{center}
\includegraphics[width=\columnwidth]{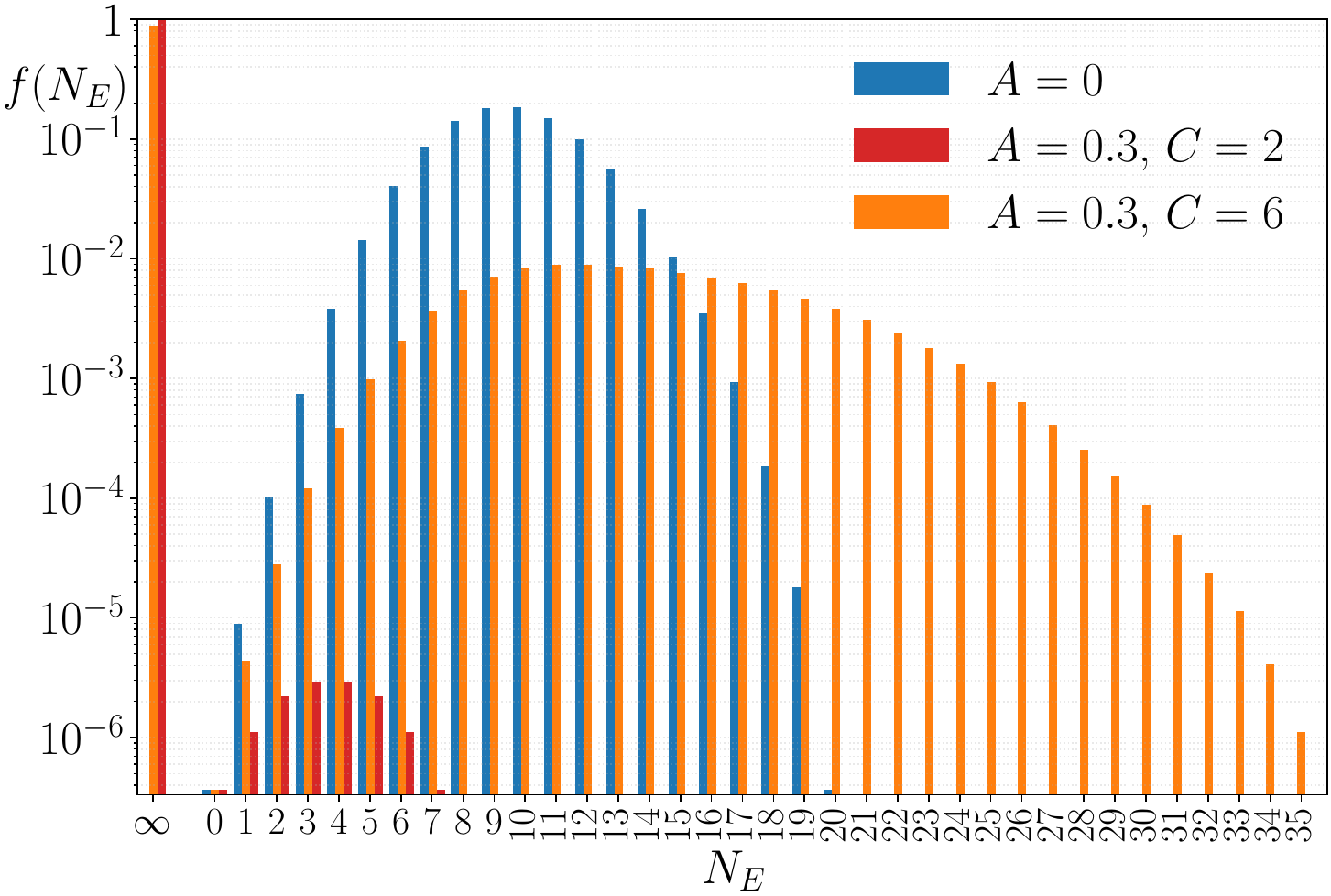}
\end{center}
\vglue -0.3cm
\caption{\label{fig1}Distribution $f(N_E)$ of quantum register Erd\H{o}s numbers $N_E$ for a quantum computer with $n_q=24$ qubits. The vertical axis gives the fraction of quantum register states having Erd\H{o}s number $N_E$. Results are shown for $A=0$ (\textcolor{myblue}{$\blacksquare$}), $A=0.3$ with $C=2$ (\textcolor{myred}{$\blacksquare$}), and $A=0.3$ with $C=6$ (\textcolor{myorange}{$\blacksquare$}). The Erd\H{o}s quantum register state (hub) is defined as the state whose energy is closest to the center of the central noninteracting band (total spin $S_z=0$). An Erd\H{o}s number (possibly infinite) is assigned to each of the $2\,704\,156$ quantum register states in the band. The column labeled $\infty$ corresponds to register states not connected to the Erd\H{o}s register state. The results are shown for one disorder realisation of Hamiltonian (\ref{hamil}).}
\end{figure}

{\it Model description.-} As a model of static QC hardware we consider $n_q$ qubits on a two-dimensional lattice with nearest-neighbor inter-qubit coupling \cite{georgeot1,georgeot2,benenti}. The system Hamiltonian introduced in \cite{georgeot1,georgeot2,benenti} is
\begin{equation}
	\label{hamil}
	H = \sum_{i} \Gamma_i \sigma_{i}^z + \sum_{i<j} J_{ij} \sigma_{i}^x \sigma_{j}^x \; ,
\end{equation}
where $\sigma_{i}$ are the Pauli matrices for qubit $i$ and the second sum runs over nearest-neighbor qubit pairs on a two-dimensional lattice with periodic boundary conditions. The energy spacing between the two states of a qubit is determined by $\Gamma_i=\Delta_0+\delta_i$, with $\delta_i$ randomly and uniformly distributed in the interval $[-\delta /2,\delta /2]$. Thus the detuning parameter $\delta$ gives the width of the $\Gamma_i$ distribution around its average value $\Delta_0$; the couplings $J_{ij}$, representing the residual interactions, are randomly and uniformly distributed in the interval $[-J,J]$. As in \cite{georgeot2,benenti}, we consider the case with $J, \delta \ll \Delta_0$ and the eigenstates in the central energy band having total conserved spin $S_z=0$. This central energy band corresponding to the highest density of states represents the QC core. The number of states in this central band is $N_b = n_q!/\left([n_q/2]!(n_q-[n_q/2])!\right)$. It was shown numerically in Refs.~\cite{georgeot1,georgeot2,benenti}
that the level-spacing statistics undergo a transition from Poisson to
Wigner--Dyson statistics above the quantum-chaos border,
\begin{equation}
	\label{qborder}
	J > J_c \approx \frac{C_c\,\delta}{n_q},
\end{equation}
with a critical parameter $C_c \approx 4$. Above this border (\ref{qborder}), the individual eigenstates reproduce the thermal Fermi-Dirac distribution (one flip of a spin implies a flip of another qubit due to total spin conservation, so that such pairs behave as fermions \cite{benenti}). The border (\ref{qborder}) follows from the {\AA}berg criterion according to which quantum chaos sets in when an interaction matrix element $H_{ij}$ is larger than the level spacing $\Delta_c$ between directly coupled states \cite{aberg1,aberg2,jacquod}. In many-body quantum systems the energy spacing between many-body states drops exponentially with the number of particles (qubits), but a majority of these states are not directly coupled due to the two-body nature of the interactions.

The relation (\ref{qborder}) is based on an average estimate of the typical spacing between directly coupled states, $\Delta_c \sim \delta/n_q$. However, it is possible to write the {\AA}berg criterion in a more exact way using numerical values of the spacings $\Delta_{ij} = \Gamma_i - \Gamma_j$ and the interaction matrix elements $J_{ij}$, which
gives
\begin{equation}
	\label{aberg}
	\mid J_{ij} \mid > A \mid \Delta_{ij} \mid \; ,
\end{equation}
where $A \sim 1$ is a certain numerical constant. In the spirit of this criterion, we can assume that interaction links that do not satisfy the condition (\ref{aberg}) are too weak and thus can be excluded from consideration (i.e., cut). For $A=0$, we have the whole network of quantum register states (e.g., $\left\vert\right\uparrow\!\uparrow\downarrow\downarrow\uparrow\dots\uparrow\rangle$) connected by interaction links, but for $A>0$ the number of links in the {\AA}berg network decreases, and we can expect that for a certain value of $A$ there will be no percolation through a global component and the network will be fragmented into small isolated clusters. Below, we describe the properties of such an {\AA}berg network for different values of the constants $A$ and $C=J\,n_q/\delta$.

{\it Properties of the {\AA}berg network.-} The frequency distribution $f(N_E)$ of the Erd\H{o}s numbers $N_E$ for a quantum register state at the center of the energy band is shown in Fig.~\ref{fig1}. For the {\AA}berg parameter $A=0$, all the network links induced by interactions $J_{ij}$ are present and the maximum of the $f(N_E)$ distribution is located at $N_E \approx 9$, which is not so far from the six degrees of separation found in the Milgram experiment \cite{milgram}. This confirms that the quantum computer hardware has a structure similar to that of small-world networks \cite{milgram,vigna,dorogovtsev,newman}. With the increase of $A$ up to $A=0.3$, certain links are cut and the connectivity of the network changes. Thus, for $A=0.3$ and $C=2$, the system is located in the phase of many-body localization \cite{georgeot2,benenti} and thus a significant number of links are cut since they do not satisfy the condition of the {\AA}berg criterion (\ref{aberg}). As a result, the network breaks up into small clusters, and a huge number of register states become disconnected from the hub, corresponding to the cell with $N_E=\infty$ in Fig.~\ref{fig1}.
In contrast, for $C=6$ the system is in the regime of quantum chaos \cite{georgeot2,benenti} and a large number of links is preserved. Thus, even if higher values of $N_E$ are required, transitions through the global network remain possible, corresponding to propagation over a whole giant component.

\begin{figure}[t]
	\begin{center}
		\includegraphics[width=\columnwidth]{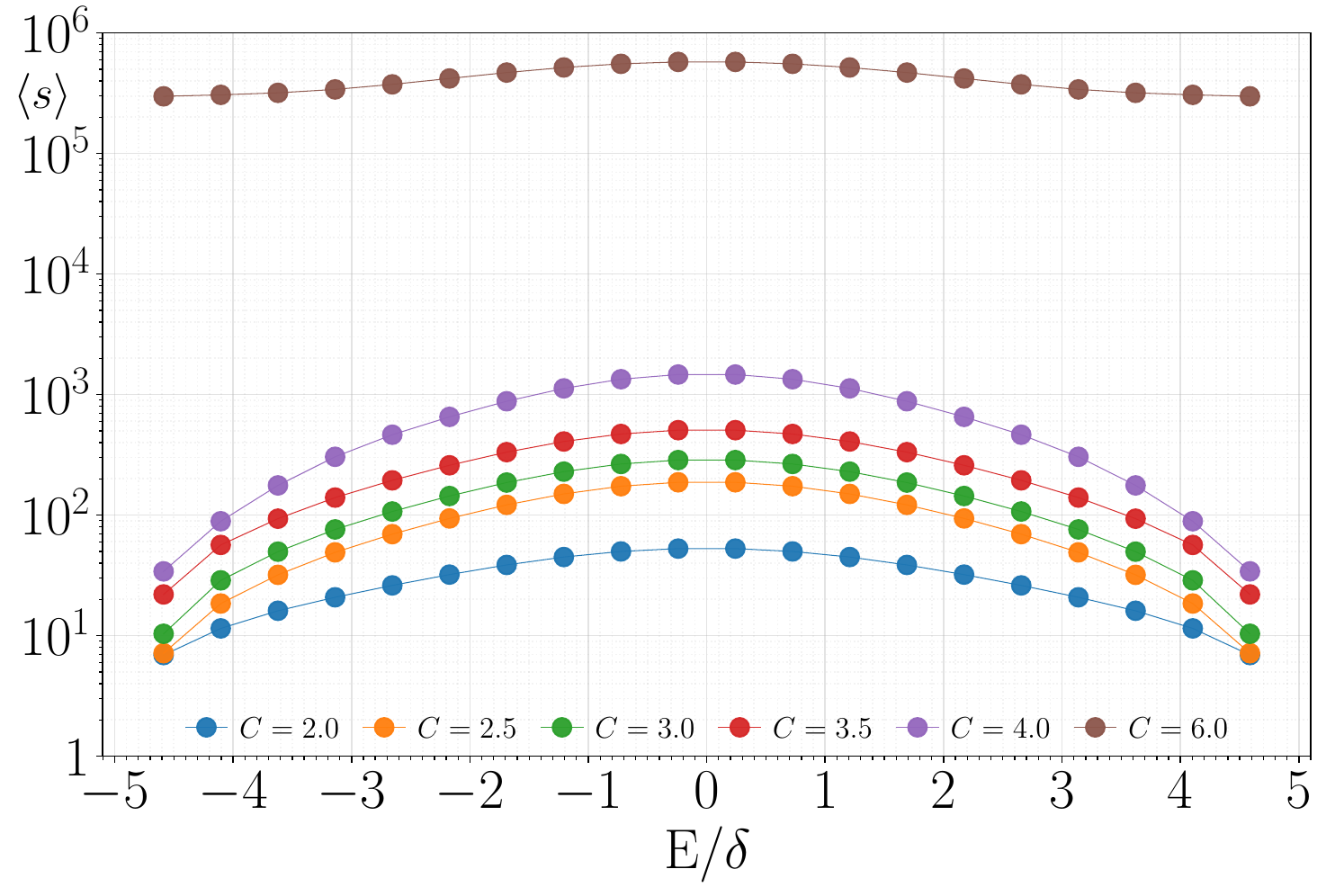}
	\end{center}
	\vglue -0.3cm
	\caption{\label{fig2}Mean reachable cluster size $\langle s\rangle$ as a function of the quantum register state energy $E$ for fixed $A=0.3$ and different interaction strengths:
		$C=2$ (\textcolor{blueprl}{$\bullet$}),
		$C=2.5$ (\textcolor{orangeprl}{$\bullet$}),
		$C=3$ (\textcolor{greenprl}{$\bullet$}),
		$C=3.5$ (\textcolor{redprl}{$\bullet$}),
		$C=4$ (\textcolor{violetprl}{$\bullet$}), and
		$C=6$ (\textcolor{brownprl}{$\bullet$}).
		The data correspond to the central noninteracting band (total spin $S_z=0$) of a quantum register network with $n_q=24$ qubits. The band is divided into 20 energy bins, and the average cluster size $\langle s\rangle$ is computed over all quantum register states belonging to each bin. The lines are drawn only as guides to the eye.}
\end{figure}

\begin{figure}[t]
	\begin{center}
		\includegraphics[width=\columnwidth]{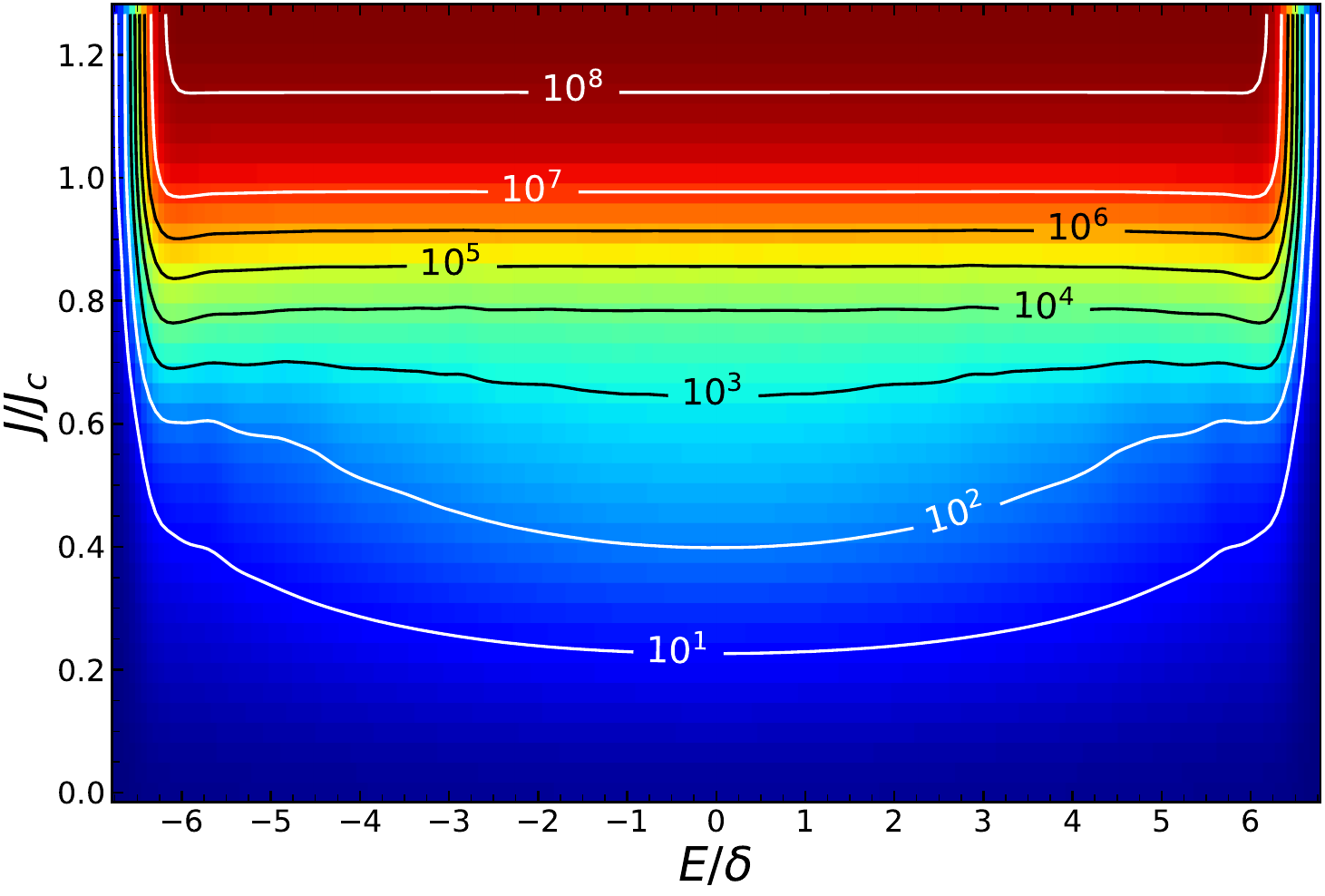}
	\end{center}
	\vglue -0.3cm
	\caption{\label{fig3}Melting of the quantum computer core for $n_q=5 \times 6 = 30$ qubits. 
		Color scale represents the logarithm of the mean reachable cluster size
		$\left\langle s\right\rangle$ varying from dark blue ($\log_{10} \left\langle s\right\rangle=0$),
		which represents clusters of size 1, to deep red ($\log_{10}\left\langle s\right\rangle\sim 8$)
		which represents clusters of size above $10^8$ (total Hilbert size is $155\,117\,520$).
		Horizontal axis is the central band energy $E/\delta$.
		Vertical axis is the value of $J/J_c$, where $J_c$ is the critical coupling $J$ for which a giant component of the {\AA}berg network contains at least half of all the quantum register states, here $J_c\simeq0.42\,\delta$. See Fig.~\ref{figS3} and Fig.~\ref{figS4} for critical values for other system sizes and the scaling of $C_c$ vs $n_q$. Results are shown for the  {\AA}berg constant $A=0.3$ and one realization of the Hamiltonian (\ref{hamil}).}
\end{figure}

The degree distributions of the {\AA}berg networks at different values of $A$ and $C$ are shown in Fig.~\ref{figS1} of the Supplemental Material (SupMat). For $A=0$, the degree of a quantum register state is equal to the number of nearest-neighbor antiparallel spin pairs in the corresponding spin configuration, or equivalently to the total number of bonds forming the domain walls of the associated square-lattice Ising configuration at fixed total spin $S_z=0$. The distributions are unimodal and remain approximately binomial over the whole parameter range, with their mean shifting toward smaller degrees as $A$ increases. They do not exhibit the broad heavy-tailed behavior characteristic of scale-free complex networks \cite{dorogovtsev,newman}.

The distributions $f(N_E)$ shown in Fig.~\ref{fig1} correspond to a hub chosen at the center of the energy band. In Fig.~\ref{fig2}, we instead let the hub vary: for a fixed {\AA}berg parameter $A=0.3$, the energy band is divided into bins, and within each bin every quantum register state is taken in turn as the hub; the resulting cluster sizes are then averaged to obtain $\langle s(E) \rangle$.
This cluster size $\langle s \rangle$ increases with increasing $C$ from $C=2$ up to $C=4$, but a very sharp increase occurs for $C=6$, corresponding to the system being in the regime of quantum chaos \cite{georgeot2,benenti}.
This shows that the {\AA}berg criterion (\ref{aberg}) allows one to correctly determine the border of many-body quantum chaos in Hamiltonian (\ref{hamil}) without expensive eigenstate computations.
The frequency distribution of the cluster sizes $\langle s\rangle$ is
shown in SupMat Fig.~\ref{figS2}.

Figure~\ref{fig3} shows, for $n_q=30$ qubits and an {\AA}berg constant $A=0.3$, the melting of the QC hardware as the residual interactions $J$ between the qubits increases. Here the critical coupling, at which 50\% of network sites belong to one cluster, is $J_c = C_c\,\delta/n_q \simeq 0.42\,\delta$, with $C_c\simeq12.6$.

\begin{figure}[t]
	\begin{center}
		\includegraphics[width=\columnwidth]{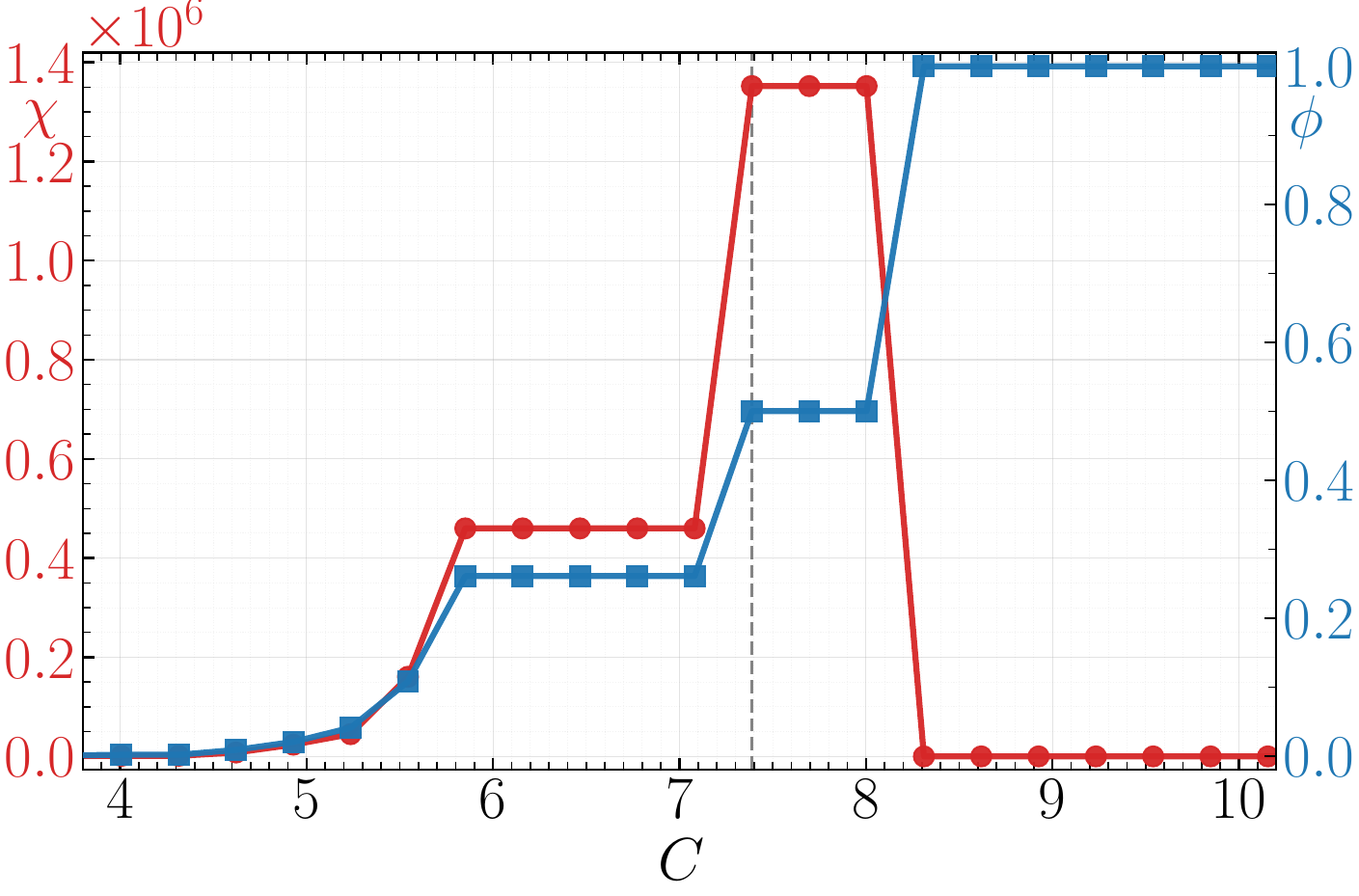}
	\end{center}
	\vglue -0.3cm
	\caption{\label{fig4}Giant connected component fraction $\phi$ and susceptibility $\chi$ as a function of the coupling constant $C$ for an \AA berg criterion $A=0.3$.
		The quantum register network corresponds to the central noninteracting band of a quantum computer with $n_q=24$ qubits. The percolation threshold is estimated as $C_c\simeq7.4$, corresponding to the smallest value of $C$ for which $\chi$ reaches its maximum.}
\end{figure}

Repeating this analysis for $n_q = 12, 16, 20, 24, 28, 30$ (see SupMat Fig.~\ref{figS3}) shows that $C_c$ grows linearly with the number of qubits, $C_c \approx B n_q$ with $B\approx 0.445$. Since the Hilbert-space size $N_b$ grows exponentially with $n_q$, this amounts to a logarithmic correction to the {\AA}berg criterion~(\ref{aberg}), $C_c \propto \log N_b$. This correction remains modest: as $n_q$ increases from 12 to 30, $N_b$ grows by four orders of magnitude ($10^4$ to $10^8$) while $C_c$ grows only by a factor of $3$.

In SupMat Fig.~\ref{figS4}, we show the QC hardware melting for $n_q =16$
($J_c\simeq0.40\,\delta, C_c\simeq6.46$) and $24$ qubits
($J_c\simeq0.31\,\delta, C_c\simeq7.44$) 
at $A=0.3$. 
In SupMat Fig.~\ref{figS5}, we compare for $n_q=16$ the obtained results with the ones obtained calculating entropy for exact diagonalization of the QC eigenstates (see Fig.~9 from \cite{georgeot2}). The comparison is done for the {\AA}berg constant $A=0.2$ giving a critical coupling value $J_c \simeq 0.27\,\delta$ $(C_c\simeq4.31)$ close to $J_c\simeq0.22\,\delta$ obtained in \cite{georgeot2} for exact eigenstates diagonalization.
Overall, the results obtained with the {\AA}berg criterion (\ref{aberg}) are similar to those obtained with exact diagonalization of the eigenstates at different interactions $J$, although visible quantitative differences remain, most pronounced near the energy band edges. Below the percolation border $J<J_c$, the {\AA}berg network still exhibits finite cluster sizes, resulting from purely classical transitions on this network (see SupMat Fig.~\ref{figS5}, left panel); by contrast, the exact quantum eigenstates (right panel) capture all the quantum effects that are absent from this classical percolation picture.
Similarities between classical percolation and quantum delocalization are known to hold for the Bethe lattice \cite{leticia}, and we find the above conclusions for percolation on the {\AA}berg network to be in qualitative agreement with those Bethe-lattice results.

\begin{figure}[t]
	\begin{center}
		\includegraphics[width=\columnwidth]{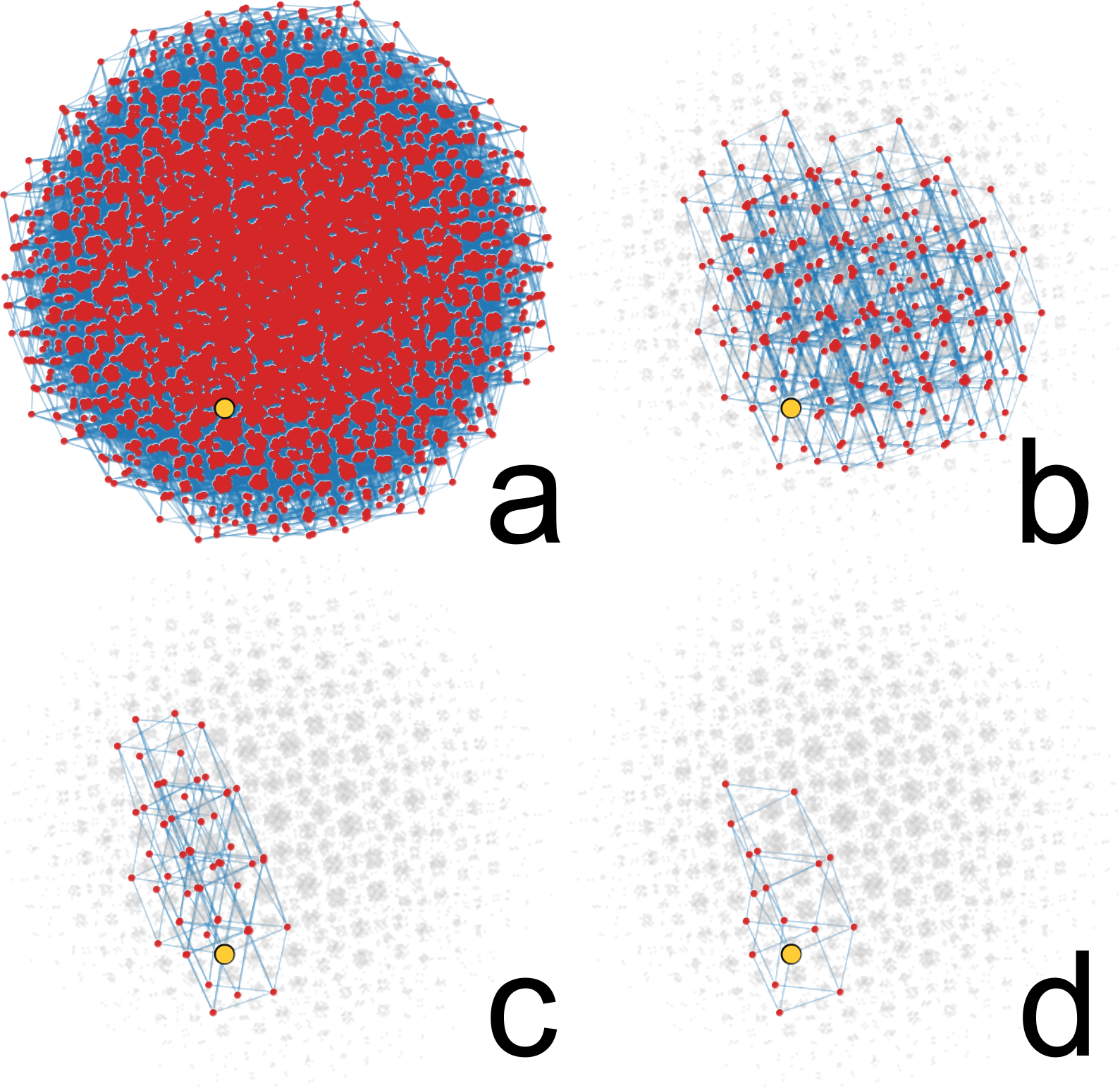}
	\end{center}
	\vglue -0.3cm
	\caption{\label{fig5}Visualization of the connected cluster containing the Erd\H{o}s quantum register (yellow node) in the quantum register network associated with the central noninteracting band of a quantum computer with $n_q=16$ qubits.
		The Erd\H{o}s quantum register state is chosen as the one whose energy is closest to zero. The coupling constant is fixed at $C=4$. Panels (a)–(d) show the evolution of this cluster as the \AA berg criterion increases $A=0.1$, $0.26$, $0.3$, and $0.5$, respectively.
		Quantum register states belonging to the connected cluster are shown as red nodes, whereas all other quantum register states are shown in gray. Interactions between quantum register states within the cluster are represented by blue edges. The corresponding cluster sizes and edge counts are $s=12\,870$, $n_\ell=68\,640$ (a); $s=336$, $n_\ell=1\,022$ (b); $s=60$, $n_\ell=142$ (c); and $s=18$, $n_\ell=33$ (d).}
\end{figure}

From Fig.~\ref{fig4}, we observe that the maximum of the susceptibility $\chi$ is found numerically to correspond to a giant connected component fraction $\phi\approx0.5$ (see mathematical definitions in SupMat), a coincidence observed for all the system sizes and parameters considered here and likely rooted in some symmetry of the {\AA}berg network. Hence, as the coupling strength increases, we define the critical coupling $C_c$, or equivalently $J_c=C_c \delta/n_q$, as the lowest value for which the susceptibility reaches its maximum and correspondingly for which the giant component of the network contains half of the quantum register states. For $n_q=24$ qubits, the maximum of the susceptibility $\chi$ is found at $C_c\approx7.4$, in agreement with the melting behavior described in Fig.~\ref{figS4} for the same system size and {\AA}berg value $A=0.3$. The susceptibility $\chi$ drops to zero as $\phi\to1$ since $\chi$ excludes the largest connected component from its computation. A plateau in $\chi$ can nonetheless be observed, since additional links may keep forming within the biggest cluster without increasing its number of nodes.

A direct visualization of the network is given in Figs.~\ref{fig5} and~\ref{figS6}, for $n_q=16$ qubits. Fig.~\ref{fig5} shows the full network of $12\,870$ nodes for different values of the {\AA}berg constant $A$, at a fixed coupling strength $C$ below the critical value for this system size; the four panels illustrate how the network shrinks as $A$ increases, i.e., as the criterion for creating or cutting links becomes more stringent. In particular, panel (c), where $A=0.3$, shows a network with few links and only a small cluster connected to the chosen hub. As $C$ is instead increased from low values toward the critical one, the network grows in size, eventually reaching a giant connected component fraction close to 1, i.e., almost all quantum register states belong to the same connected component. This is confirmed in Fig.~\ref{figS6}, where $A$ is fixed at $0.3$ and $C$ is varied, showing that the network becomes nearly fully connected close to the critical value; see panel (d).

{\it Discussion.-} We demonstrated that the network of quantum register states, whose links stem from the residual interactions between qubits of a quantum computer, has the small-world properties of complex networks found in human society. The distribution of Erd\H{o}s numbers shows that any quantum register state can be reached from another via about 9 interaction links (see Fig.~\ref{fig1}), close to the six degrees of separation found by Milgram between people in the US \cite{milgram}.
We analyzed the properties of this QC network using the {\AA}berg criterion (\ref{aberg}) \cite{aberg1,aberg2}, according to which a coupling is effective only if it exceeds the energy spacing between the coupled states. Retaining only the links satisfying this criterion yields the {\AA}berg network, whose number of links increases with the interaction strength. We showed that above a certain critical interaction $J_c$, the {\AA}berg network percolates, with the giant component covering practically the whole QC network; below this threshold ($J<J_c$), only localized clusters remain, with decreasing size as $J$ is reduced.
This {\AA}berg-criterion approach gives access to system sizes up to $n_q=30$ qubits, well beyond what is reachable by exact diagonalization, while remaining in good agreement with the latter where a direct comparison is possible. We argue that it offers a new perspective on the properties of QC interaction networks.

\noindent {\it Acknowledgments.-}
This work was partially supported by the NANOX project (Grant No.\ ANR-17-EURE-0009)
within the Programme d'Investissements d'Avenir (project MTDINA).
It was also supported by the Compétences et
Métiers d'Avenir France 2030 QuanTEdu-France project (Grant No.\ ANR-22-CMAS-0001),
the EIPHI Graduate School (Grant No.\ ANR-17-EURE-0002), and
the Bourgogne--Franche-Comté Region (projects ADN and CCNV).

\noindent {\it Data availability.-} Data are available from the authors upon reasonable request.

\pagebreak


\clearpage
\onecolumngrid

\appendix

\renewcommand{\thesection}{S\arabic{section}}
\renewcommand{\theequation}{S\arabic{equation}}
\renewcommand{\thefigure}{S\arabic{figure}}
\renewcommand{\thetable}{S\arabic{table}}

\setcounter{section}{0}
\setcounter{equation}{0}
\setcounter{figure}{0}
\setcounter{table}{0}

\noindent{\Large Supplemental Material for \\{\bf Small-world structure of quantum computer hardware}}

\noindent by S. J. da Silva Junior,$^1$ D. L. Shepelyansky,$^2$  and J. Lages$^1$

\noindent $^1$Université Marie et Louis Pasteur, CNRS, Institut UTINAM, équipe de physique théorique, Besançon, France

\noindent $^2$Univ Toulouse, CNRS, Laboratoire de Physique Théorique, Toulouse, France\\

\noindent August 13, 2026

\section*{Mathematical definitions}

Let us consider a network of size $N$. We define the mean reachable cluster size as
\begin{equation}
	\left\langle s \right\rangle
	=
	\frac{\sum_{s=1}^{N}s^{2}n_s}
	{\sum_{s=1}^{N}s\,n_s},
\end{equation}
where the sum runs over all possible cluster sizes $s$, and $n_s$ denotes the number of connected components of size $s$. The quantity $\langle s\rangle$, which ranges from $1$ to $N$, measures the average size of the connected component reached by selecting a node uniformly at random, and therefore characterizes the effective connectivity of the network from a local perspective.

The giant connected component fraction is defined as
\begin{equation}
	\phi=\frac{s_{\max}}{N},
\end{equation}
where
\begin{equation}
	s_{\max}=\max\{\,s:\,n_s>0\,\}
\end{equation}
is the size of the largest connected component.

The susceptibility is defined as
\begin{equation}
	\chi=
	\frac{\sum_{s<s_{\max}}s^{2}n_s}
	{\sum_{s=1}^{N}s\,n_s},
\end{equation}
where the numerator excludes the largest connected component. The susceptibility satisfies
$0\le\chi\le N\phi(1-n_{s_{\max}}\phi)$ where $n_{s_{\max}}$ denotes the number of connected components of maximum size.
It vanishes when all nodes are isolated or when the entire network forms a single connected component, and reaches a maximum in the vicinity of the percolation transition.

The three quantities are related through
\begin{equation}
	\langle s\rangle
	=
	N\,n_{s_{\max}}\phi^{2}
	+\chi.
\end{equation}

\section{Degree distributions of the {\AA}berg network}

\begin{figure}[t]
	\begin{center}
		\includegraphics[width=\columnwidth]{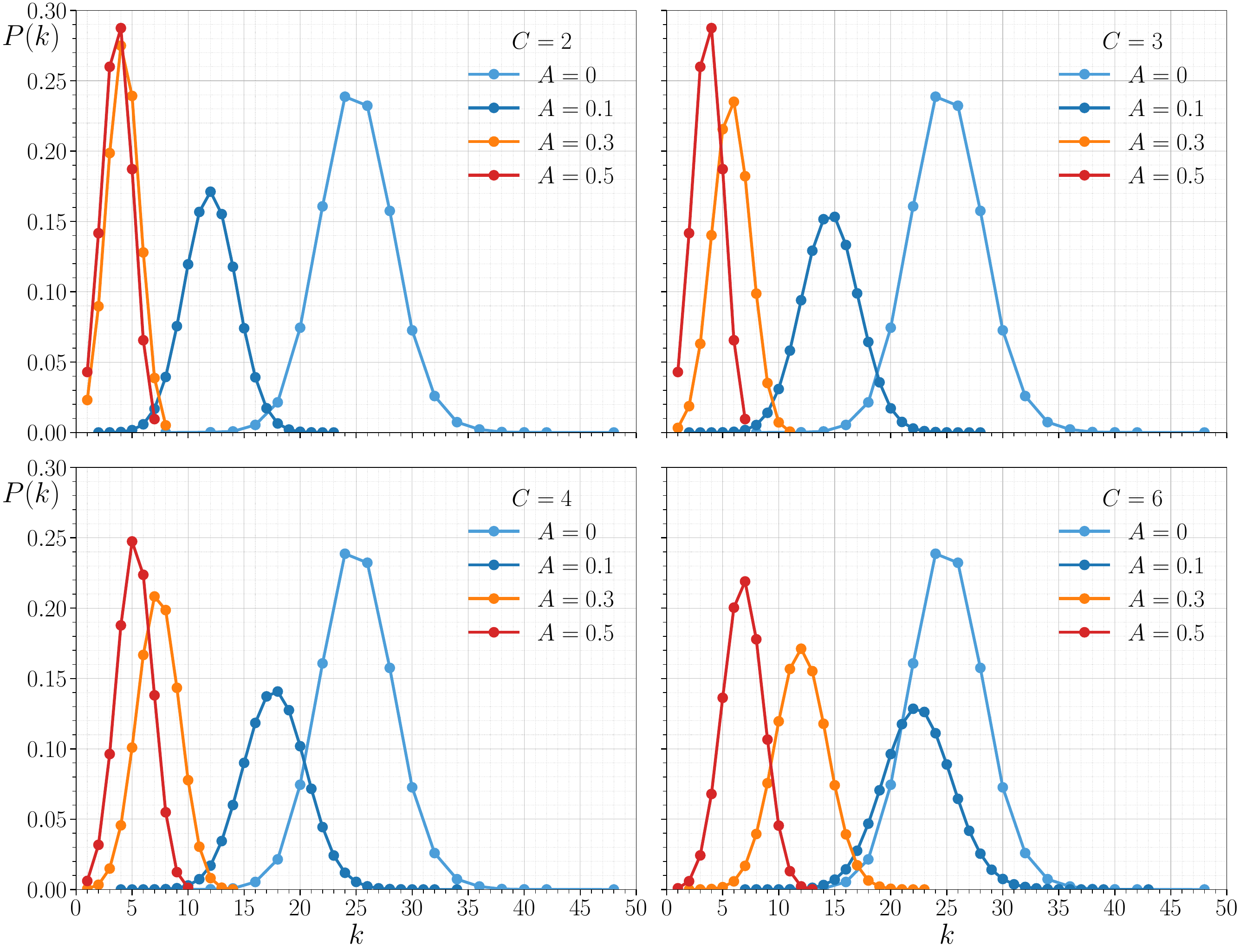}
	\end{center}
	\vglue -0.3cm
\caption{\label{figS1}Degree distribution $P(k)$ of the {\AA}berg networks for a quantum
	computer hardware with $n_q=24$ qubits, where $k$ denotes the number of
	links incident on a quantum-register node.
	The cases $C=2$ (top left), $C=3$ (top right), $C=4$ (bottom left), and
	$C=6$ (bottom right) are shown for
	$A=0$ (\textcolor{colorA0}{$\bullet$}),
	$A=0.1$ (\textcolor{colorA01}{$\bullet$}),
	$A=0.3$ (\textcolor{colorA03}{$\bullet$}), and
	$A=0.5$ (\textcolor{colorA05}{$\bullet$}).
	The distributions are computed over the $2\,704\,156$ quantum register states
	belonging to the central noninteracting band with total spin $S_z=0$.
	The results correspond to a single realization of Hamiltonian~(\ref{hamil}).}
\end{figure}

Figure~\ref{figS1} shows the degree distributions $P(k)$ of the
{\AA}berg networks for different values of the {\AA}berg parameter $A$ and coupling strength $C$, where $k$ denotes the degree of a quantum register state in the corresponding network.

For $A=0$, all nearest-neighbor interaction bonds are retained.
The degree of a quantum register state is therefore determined solely by its spin configuration. Each nearest-neighbor antiparallel spin pair generates one transition through the
$\sigma_i^x\sigma_j^x$ interaction, so that the degree is equal to the number of nearest-neighbor antiparallel spin pairs, or equivalently to the total number of bonds forming the domain walls of the associated square-lattice Ising configuration at fixed total spin $S_z=0$.
Consequently, only even values of the degree occur, see $A=0$ cases in Fig.~\ref{figS1}. The complete square-lattice geometry induces strong correlations between neighboring bonds, resulting in noticeable deviations from a binomial distribution.

For $A>0$, the {\AA}berg criterion removes interaction bonds according to relation (\ref{aberg}), producing diluted interaction networks.
Remarkably, the resulting degree distributions are found to be very accurately described by effective binomial laws,
\begin{equation}
P(k)\approx
\binom{M}{k}
p^k(1-p)^{M-k},
\end{equation}
where $M$ and $p$ are determined from the first two moments of the degree distribution.

The fitted probability is found to be nearly independent of both
$A$ and $C$,
and is accurately described by
\begin{equation}
p=\frac{n_q}{2(n_q-1)},
\end{equation}
where $n_q$ is the number of qubits.
This expression has a simple combinatorial interpretation. Indeed, in the central noninteracting band, all spin configurations with $N_\uparrow=N_\downarrow=n_q/2$ are equiprobable.
For any nearest-neighbor bond $\langle i,j\rangle$,
\begin{equation}
\begin{aligned}
	P(\tau_i\neq\tau_j)
	&=
	P(\uparrow_i,\downarrow_j)
	+
	P(\downarrow_i,\uparrow_j)\\
	&=
	\frac12\frac{n_q/2}{n_q-1}
	+
	\frac12\frac{n_q/2}{n_q-1}
	=
	\frac{n_q}{2(n_q-1)},
\end{aligned}
\end{equation}
which is therefore the probability that a given interaction bond connects two antiparallel spins.
For the present system with $n_q=24$ qubits (Fig.~\ref{figS1}) this gives
\begin{equation}
p=\frac{12}{23},
\end{equation}
while in the thermodynamic limit
\begin{equation}
p\rightarrow\frac12.
\end{equation}

For $A>0$, the agreement between the numerical degree distributions and the
effective binomial laws is excellent, with typical total variation
distances below $3\times10^{-2}$ and Kullback--Leibler divergences
below $10^{-3}$.

\section{Additional figures}

\begin{figure}[h]
	\includegraphics[width=\columnwidth]{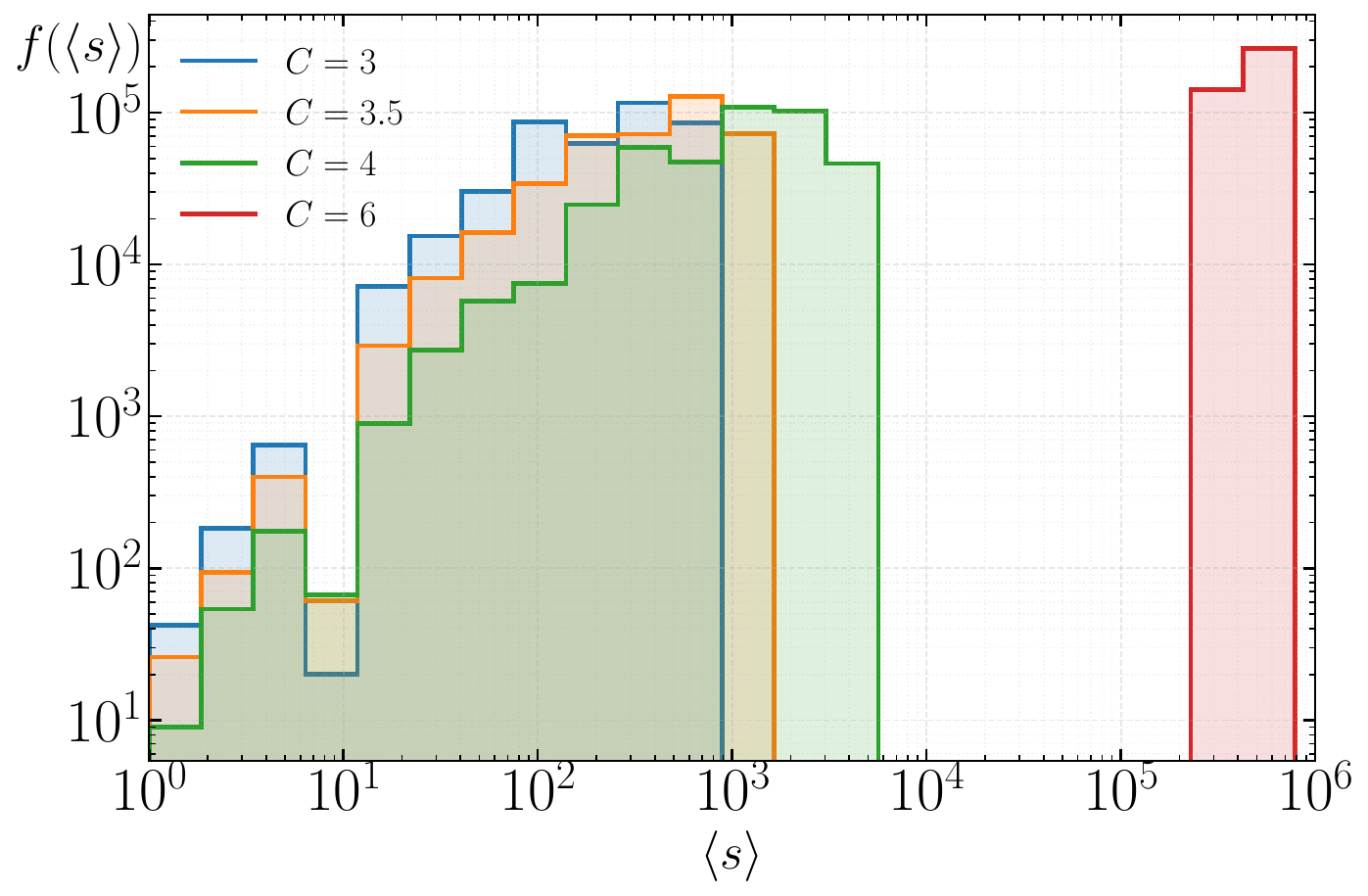}
	\caption{\label{figS2}Histogram $f\left(\left\langle s\right\rangle\right)$ of the mean reachable cluster size $\left\langle s\right\rangle$ for quantum register states whose energies belong to the interval $E\in[-0.48,0)$ in the central noninteracting band of a quantum computer hardware with $n_q=24$ qubits. The horizontal axis is divided into 24 logarithmically spaced bins in $\left\langle s\right\rangle$, and the vertical axis gives the number $n(\left\langle s\right\rangle)$ of quantum register states falling into each bin. Results are shown for $A=0.3$ and $C=3$ (\textcolor{blueprl}{$\bullet$}), $C=3.5$ (\textcolor{orangeprl}{$\bullet$}), $C=4$ (\textcolor{greenprl}{$\bullet$}), and $C=6$ (\textcolor{redprl}{$\bullet$}).}
\end{figure}

\begin{figure}[h]
	\begin{center}
		\includegraphics[width=\columnwidth]{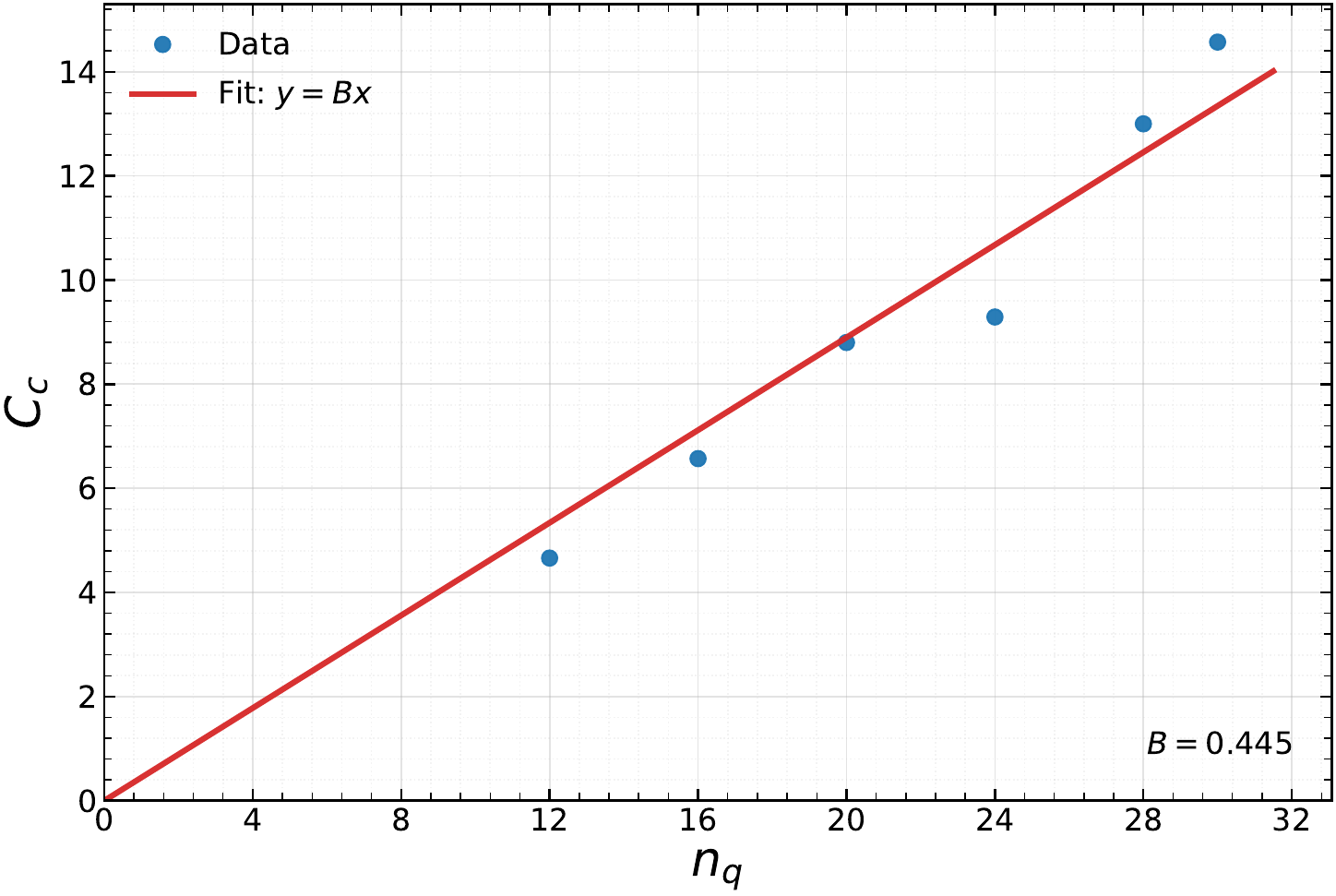}
	\end{center}
	\vglue -0.3cm
	\caption{\label{figS3} Linear fit of the critical coupling $C_c$, defined as the coupling for which there is a subset of the network that contains $50\%$ of all the states, against the number of qubits $n_q=12, 16, 20, 24, 28$ and $30$. The critical values $C_c$ are obtained by
		an average over several disorder realisations. The fit shows a linear dependence of the form $C_c = Bn_q$ with $B\simeq0.445$.}
\end{figure}

\begin{figure}
	\includegraphics[width=\columnwidth]{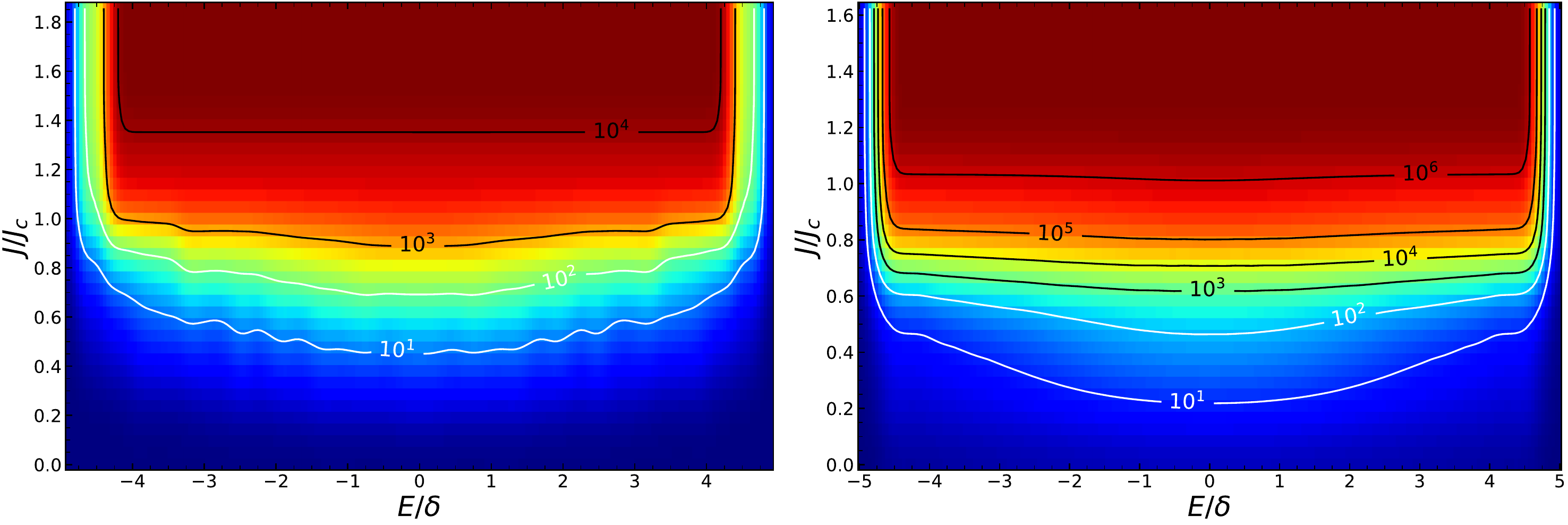}
	\caption{\label{figS4}Melting of the quantum computer core for two systems of $n_q=16$ (left panel) and $24$  (right panel) qubits. Color scale represents the logarithm of the mean reachable cluster size $\left\langle s\right\rangle$, varying from dark blue ($\log_{10} \left\langle s\right\rangle=0$), which represents clusters of size 1, to deep red ($\log_{10}\left\langle s\right\rangle\sim 4$, left panel, and $\log_{10}\left\langle s\right\rangle\sim 6$, right panel), which represents clusters of sizes above $10^4$ and $10^6$ for $n_q=16$ and $n_q=24$, respectively.   Horizontal axis is the central band energy $E/\delta$. Vertical axis is the value of $J/J_c$ with $J_c\simeq0.40\,\delta$ (left panel) and $J_c\simeq0.31\,\delta$ (right panel). Results are shown for $A=0.3$ and one realization of the Hamiltonian (\ref{hamil}).}
\end{figure}

\begin{figure}
	\centering
    \includegraphics[width=\columnwidth]{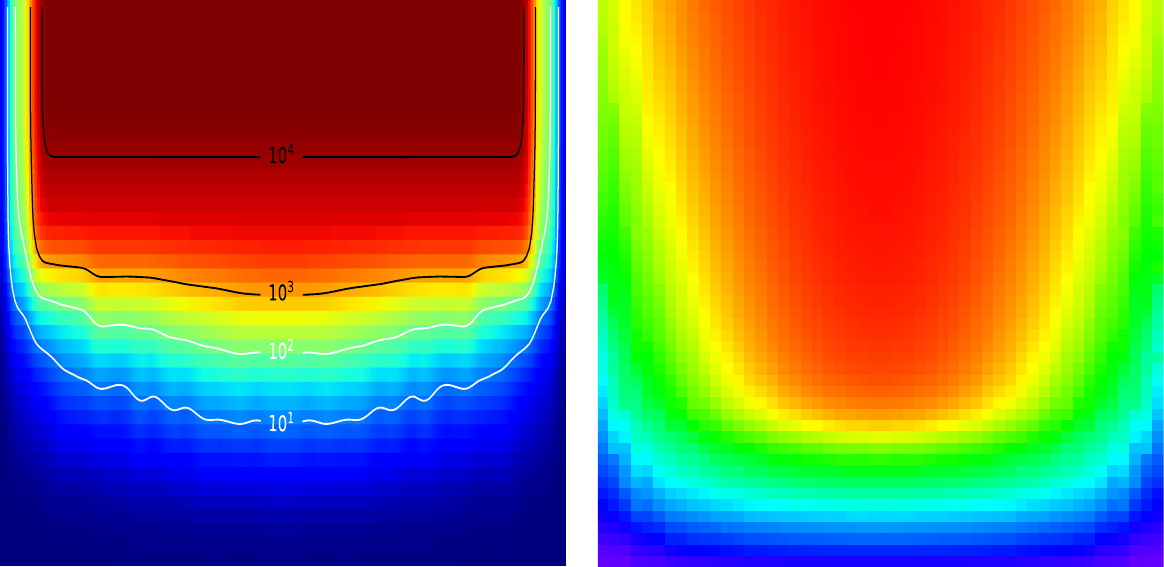}	
	\caption{\label{figS5}Melting of the quantum computer core for $n_q=16$ qubits. Left panel: color scale represents the logarithm of the mean reachable cluster size $\left\langle s\right\rangle$, varying from dark blue ($\log_{10} \left\langle s\right\rangle=0$), which represents clusters of size 1, to deep red ($\log_{10}\left\langle s\right\rangle\sim 4$), which represents clusters of size above $10^4$. Right panel: Fig.~9 from \cite{georgeot2}, color represents the level of the quantum eigenstate entropy $S_q$, from bright red ($S_q$ $\approx$ 12) to blue ($S_q = 0$); see details in \cite{georgeot2}. Horizontal axis is the central energy band $E/\delta$. Vertical axis is the value of $J$, varying from $0$ to $0.5$ in both left and right panels. Results are shown for $A=0.2$ (left panel), and one realization of the Hamiltonian (\ref{hamil}).}
\end{figure}

\begin{figure}[h]
	\begin{center}
		\includegraphics[width=\columnwidth]{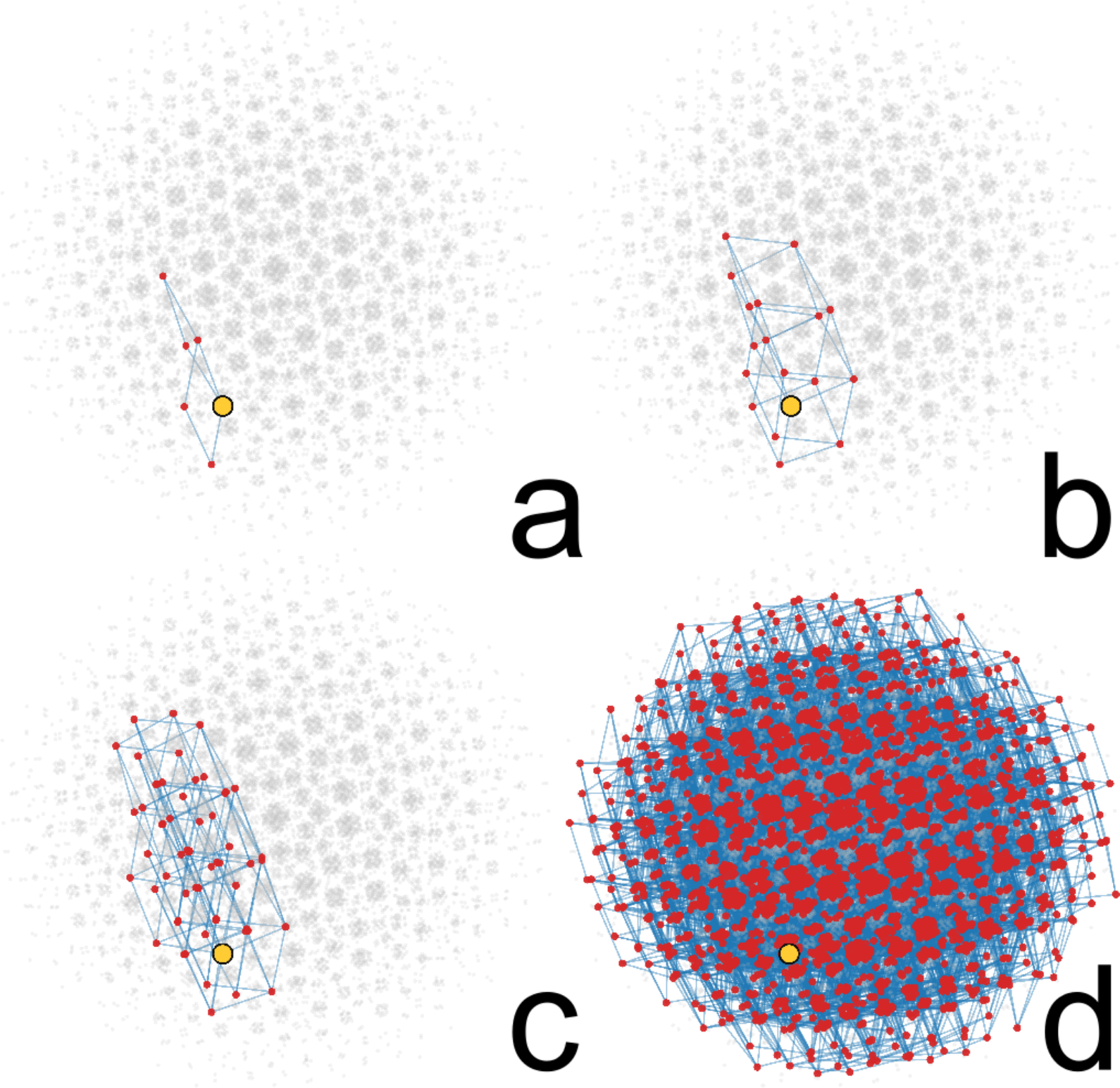}
	\end{center}
	\vglue -0.3cm
	\caption{\label{figS6}Visualization of the connected cluster containing the Erd\H{o}s quantum register state (yellow node) in the quantum register network associated with the central noninteracting band of a quantum computer with $n_q=16$ qubits. The Erd\H{o}s quantum register state is chosen as the one whose energy is closest to zero. The \AA berg criterion is fixed at $A=0.3$. Panels (a)–(d) show the evolution of this cluster as the coupling constant increases $C=2$, $3$, $4$, and $6$, respectively. Quantum register states belonging to the connected cluster are shown as red nodes, whereas all other quantum register states are shown in gray. Interactions between quantum register states within the cluster are represented by blue edges. The corresponding cluster sizes and edge counts are $s=6$, $n_\ell=7$ (a); $s=18$, $n_\ell=33$ (b); $s=60$, $n_\ell=142$ (c); and $s=3\,432$, $n_\ell=12\,804$ (d).}
\end{figure}

\end{document}